# On the Gravitational Energy of the Hawking Wormhole

H.Culetu[*]

Ovidius University, Dept.of Physics,

B-dul Mamaia 124, 8700 Constanta, Romania

**Abstract**

The surface energy for a conformally flat spacetime which represents the Hawking wormhole in spherical (static) Rindler coordinates is computed using the Hawking – Hunter formalism for non asymptotically – flat spacetimes. The physical gravitational Hamiltonian is proportional to the Rindler acceleration g of the hyperbolic observer and is finite on the event horizon $\xi = b$ (b-the Planck length, $\xi$- the Minkowski interval). The corresponding temperature of the system of particles associated to the massless scalar field $\Psi = 1 - b^2/\xi^2$, coupled conformally to Einstein's equations, is given by the Davies - Unruh temperature up to a constant factor of order unity.

**1) Introduction**

There has recently been a special interest in the Hamiltonian formulation of General Relativity motivated by including more general boundary conditions[1-3]. The surface term that arises in the Hamiltonian can be taken as the definition of the total energy, even for spacetimes that are not asymptotically flat (AF) (in the AF case, it agrees with the usual ADM energy).

---

*e-mail: hculetu@yahoo.com

Casadio and Gruppuso[4] analyzed the behaviour of the total energy in AF spacetimes as it results from surface terms, reaching the conclusion that no boundary terms are generated by a conformal transformation in Lagrangean formalism, but the ADM mass is changed in Hamiltonian formalism, when the conformal factor contains terms of order $r^{-1}$.

Garattini[5] defined a quasilocal energy as the value of the Hamiltonian that generates time translations orthogonal to the two dimensional boundaries (each of which is located in one of the two causally disconnected regions), an expression which has formally a Casimir form.

The porpose of the paper is to compute the gravitational energy using the Hawking – Hunter prescription[3] for the (Lorentzian version) of the Hawking wormhole[6]. We chose that spacetime as it is conformally - flat (with a Lorentz-invariant conformal factor) and the metric does not depend on time, in spherical Rindler coordinates. In addition, the Hawking wormhole geometry is different compared to Minkowski geometry only very close to the light-cone[7]. We have also the possibility to relate the gravitational energy to the Davies – Unruh temperature since a spherical Rindler observer undergoes a uniformly accelerated motion.

The Hawking wormhole is a solution of the conformally invariant Einstein equations coupled to a massless conformally scalar field[7].

In Sec. 2 we introduce the Hawking wormhole geometry in a static form, with the help of the Gerlach version[8] of the Rindler spacetime in spherical coordinates.

Using the Hawking – Hunter[3] prescription, we compute in Sec.3 the physical Hamiltonian (the total energy) for an observer fixed at $\xi$ = const. in the accelerated system). On the horizon $\xi$ = b, the surface energy of the (Planck sized) bubble is proportional to the acceleration g, similar to the thermal energy (per particle) from the Davies – Unruh formula for the temperature of the thermal bath[9-11].

We could consider these Planck – sized bubbles as „particles" associated to the massless conformal scalar field coupled to Einstein equations and the energy per particle would be kT ~ g, k being the Boltzmann constant and g – the modulus of the constant acceleration measured by an inertial observer momentarily at rest with respect to the accelerated one.

From now on the units will be such that $c = G = k = \hbar = 1$, excepting special cases.

## 2. The spherical Rindler frame

In order to avoid the special direction given by acceleration, we study the expression of the surface energy in spherical Rindler coordinates. We chose the static form of the Rindler geometry in order to have the possibility to define energy. Therefore, we make use of the coordinate transformation

$$x^1 = \xi \sin\theta \cos\varphi, \qquad x^2 = \xi \sin\theta \sin\varphi,$$
$$x^3 = \xi \cos\theta \cosh gt, \qquad x^0 = \xi \cos\theta \sinh gt, \qquad (2.1)$$

By means of (2.1) the Minkowski spacetime becomes

$$ds^2 = -g^2 \xi^2 \cos^2\theta \, dt^2 + d\xi^2 + \xi^2 \, d\Omega^2 \qquad (2.2)$$

which has the desired static form.

It is a well known result that an accelerated observer detects particles as if it were in a thermal heat bath at the temperature $(g/2\pi)$[9,10]. Why an Unruh detector click when it is accelerated, even in Minkowski space ? Basing on the fact that the v.e.v $<T_{\mu\nu}>$ is a covariant object, the answer would be „no". If the regularized stress tensor $<T_{\mu\nu}>$ vanishes in one frame (say the Minkowski frame), than it must vanish in all frames and hence in the Rindler frame. On the other hand, the detector would not have to click as long as it will never see any curvature of the spacetime.

We propose to avoid the contradiction by replacing the flat space with a conformally flat one, the conformal factor being Lorentz – invariant.

Therefore we compute the Gravitational Hamiltonian for the Lorentzian version of the Hawking wormhole, which is Minkowski spacetime far from the light cone[6]

$$ds^2 = \left(1 - \frac{b^2}{x_\alpha x^\alpha}\right)^2 \eta_{\mu\nu} \, dx^\mu \, dx^\nu \tag{2.3}$$

where „b" is the „neck radius" of the wormhole (which will be taken of the order of the Planck length), $\eta_{\mu\nu} = \text{diag}(-1,1,1,1,)$ and $x_\alpha x^\alpha \; (\alpha = 0,1,2,3)$ is the square of the Minkowski interval.

The transformation (2.1) changes the geometry (2.3) into

$$ds^2 = \left(1 - \frac{b^2}{\xi^2}\right)^2 \left(-g^2 \xi^2 \cos^2\theta \, dt^2 + d\xi^2 + \xi^2 \, d\Omega^2\right) \tag{2.4}$$

We have horizons at $\xi = b$ and $\theta = \pi/2$ (because of the conformal factor, $\xi = 0$ is no longer a horizon, as in the Rindler geometry).

Only the region $\xi \geq b$ will be studied in this paper.

### 3. The surface energy of the accelerated observer

As it was stated in Introduction, with the help of accelerated coordinates we have the possibility of finding a connection between the surface energy and the Davies – Unruh effect.

Using the Hawking – Hunter prescription[3], we compute the Gravitational Hamiltonian for the spacetime (2.4).

As Casadio and Grupusso[4] have shown, no boundary terms are generated by a conformal transformation in the Lagrangean formalism. Since in our case the conformal factor $(1-b^2/\xi^2)^2$ depends quadratically on the radial coordinate $\xi$, no boundary terms are generated by a conformal transformation in the Hamiltonian formalism also. Therefore, „the constrained term" [3] vanishes.

Since, as we shall see later, the unit normals $n^\alpha$ and $u^\alpha$ to the spacelike hypersurface $\Sigma_t$ of constant t and to the timelike boundary B of constant $\xi$, respectively, are orthogonal, „the tilting term" in the Hamiltonian also vanishes. „The momentum term" is vanishing as we are dealing with a diagonal metric (the shift vector is zero).

We therefore remain with only one term in the Hamiltonian, „the curvature term"

$$H = -\frac{1}{8\pi} \int_{B_t} N \sqrt{\sigma} K \, d^2x, \qquad (3.1)$$

where $B_t = \Sigma_t \cap B$, N – the lapse function, $\sigma$ - the determinant of the metric on $B_t$ and K – the corresponding trace of the extrinsic curvature.

Basing on the method developed by Berezin, Kuzmin and Tkachev [12], we find the metric on $\Sigma_t$ to be

$$h_{ij} = \left(1 - \frac{b^2}{\xi^2}\right)^2 \left(1, \xi^2, \xi^2 \sin^2\theta\right), \quad (i,j=1,2,3) \qquad (3.2)$$

while $N = g\xi(1 - b^2/\xi^2)\cos\theta$. The unit normal to $\Sigma_t$ is

$n_\alpha = \left[g\xi(1-b^2/\xi^2)\cos\theta, 0, 0, 0\right]$

with $n_\alpha n^\alpha = -1$ and the unit normal to B appears as

$u_\alpha = \left(0, 1 - b^2/\xi^2, 0, 0\right), \qquad (u^\alpha u_\alpha = 1)$

The metric on $B_t$ can be written as

$\sigma_{\mu\nu} = h_{\mu\nu} - u_\mu u_\nu = \left[0, 0, \xi^2(1-b^2/\xi^2)^2, \xi^2(1-b^2/\xi^2)^2 \sin^2\theta\right]$

with $\sqrt{\sigma} = \xi^2(1-b^2/\xi^2)^2 \sin\theta$ and $h_{\mu\nu} = g_{\mu\nu} + n_\mu n_\nu$.

The geometry on the surface $\xi = \xi_0$ is

$$\gamma_{\mu\nu} = g_{\mu\nu} - u_\mu u_\nu = \left(1 - \frac{b^2}{\xi^2}\right)^2 \left(-g^2 \xi^2 \cos^2\theta, 0, \xi^2, \xi^2 \sin^2\theta\right) \tag{3.3}$$

and its extrinsic curvature looks as

$$\Theta_{\mu\nu} = \gamma_\mu^\alpha \nabla_\alpha u_\nu = \nabla_\mu u_\nu - u_\mu u^\alpha \nabla_\alpha u_\nu.$$

Hence, the trace of the extrinsic curvature of $B_t$ acquires the form

$$K = \Theta - u_\mu a^\mu$$

where $a^\mu = n^\nu \nabla_\nu n^\mu$ is the acceleration of the unit normal $n^\mu$. We have, in our geometry (2.4)

$$\Theta = \frac{1}{\xi}\left(1 + \frac{b^2}{\xi^2}\right)\left(1 - \frac{b^2}{\xi^2}\right)^{-2}\left(2 + \frac{1}{\cos^2\theta}\right)$$

Keeping in mind the components of $\gamma_{\mu\nu}$ from (3.3), we obtain

$$K = \frac{2}{\xi}\left(1 + \frac{b^2}{\xi^2}\right)\left(1 - \frac{b^2}{\xi^2}\right)^{-2}, \text{ at } \xi = \xi_0. \tag{3.4}$$

We are now in a position to write down the value of H from (3.1)

$$H = -\frac{g}{4}\xi_0^2\left(1 - \frac{b^4}{\xi_0^4}\right) \tag{3.5}$$

In order to get the physical Hamiltonian $H_P$, we have to subtract the Minkowski contribution, namely, the contribution at infinity ($\xi_0 \gg b$)

$$H_P = \frac{gb^2}{4\xi_0^2} \tag{3.6}$$

where we took advantage of the fact that $b^2 = l_P^2 = G$.

But we know that we have in fact two Rindler observers (one in each hemisphere, $0 \leq \theta < \pi/2$ and $\pi/2 < \theta \leq \pi$), which are causally disconnected. An integration over θ in the second hemisphere changes the sign of $H_P$. A similar result was obtained by Garattini[5]. We have finally

$$\tilde{H}_P = 2H_P = \frac{g b^2}{2\xi_0^2} \tag{3.7}$$

The spacetime (2.4) for which we computed the Gravitational Hamiltonian has two horizons at θ = π/2 and ξ = b. The hypersurface ξ = b is the Hawking wormhole which separates the two (causally disconnected) AF regions with positive and, respectively, negative energy.

For the special value ξ = b, (3.7) becomes

$$\tilde{H}_P = \frac{g}{2} \tag{3.8}$$

Eq. (3.7) represents the total gravitational energy on the surface ξ = $\xi_0$. On the horizon we have $\xi_0$ = b. We might take all „bubbles" of radius b as being „particles" associated to the scalar field ψ = 1- $b^2/\xi^2$, coupled conformally to the Einstein equations. Keeping in mind that the massless field Ψ is Lorentz-invariant (it depends only on ξ = $x^\alpha x_\alpha$), the particles of the field (bubbles of radius b) have no a proper reference system. In addition, their Planck length radius is also an invariant (ξ = b is a null surface, where the energy of the particle is localized). Taking the thermal energy of the particles to be (3/2) kT, eq. (3.18) yields

$$T = \frac{g}{3} \tag{3.9}$$

an expression which approximates very well the Davies - Unruh formula for the temperature of the thermal bath in which a Rindler observer is immersed.

When (3.8) is written under the form

$$\tilde{H}_P = \frac{\hbar c}{2\frac{c^2}{g}},$$

it resembles the Casimir energy between two (perfectly conducting) plates. In our case, we have a „spherical box" with a diameter $2c^2/g = \xi$, the wall of the box playing the same role as the perfectly reflecting plates.

## 4. Conclusions

We have studied in this paper the gravitational energy as a surface integral in a geometry conformal to the Rindler spacetime. This geometry (the Hawking wormhole) is a solution of Einstein's equations coupled conformally to a massless scalar field $\Psi$.

We found the physical Hamiltonian $\tilde{H}_P$ is proportional to the acceleration of the Rindler observer and the Planck length b squared. The corresponding temperature of the system of „particles" (bubbles) of radius b and associated to the conformally scalar field seems to be simply the Davies – Unruh temperature.

The energy of the particles is localized on the null surface $\xi = b$ and the bubbles have no a proper reference system because the massless field $\Psi$ is Lorentz-invariant.

## Acknowledgements
I would like to thank Roberto Casadio for some useful comments.I would like to thank Roberto Casadio for some useful comments.